\documentclass[english,prl,twocolumn,groupedaddress,reprint]{revtex4}
\usepackage{amsfonts}
\usepackage[T1]{fontenc}
\usepackage[latin9]{inputenc}
\usepackage{color}
\usepackage{amsmath}
\usepackage{amssymb}
\usepackage{graphicx}
\usepackage{esint}
\usepackage{comment}
\usepackage{bbold}
\usepackage{tikz}
\usepackage{babel}
\usepackage{minitoc}
\usepackage{bm}
\usepackage[caption=false]{subfig}

\begin{document}

\title{Precision metrology of weak measurement with thermal state pointer}
\author{Gang Li\footnote{ligang0311@sina.cn},$^{1}$ Li-Bo Chen,$^{2}$ Tao Wang\footnote{suiyueqiaoqiao@163.com},$^{3}$ Zhi-Hui He,$^{1}$ and He-Shan Song$^{4}$}
\affiliation{$^{1}$School of Physics and Electronic Information, Yan'an University, Yanan
716000, China }
\affiliation{$^{2}$School of Science, Qingdao Technological University, Qingdao 266033, China}
\affiliation{$^{3}$College of Physics, Tonghua Normal University, Tonghua 134000, China}
\affiliation{$^{4}$School of Physics, Dalian University Of Technology, Dalian 116024, China}
\date{\today }

\begin{abstract}
Quantum metrology is being gradually studied for weak measurement systems.
For weak measurement systems with thermal state pointer, we find that in the
displacement space corresponding to imaginary weak values, the maximal QFI
after successful postselection can attain the level of thermal fluctuations,
without surpassing total QFI, and that QFI which increases with increasing
temperature can constantly improve the measurement precision. These results
are much better than that of weak measurement with pure state (i.e.,
Gaussian state) pointer. On the other hand, in Kerr nonlinear interaction
systems with weak measurement, and by using thermal state pointer, we obtain
in the phase space successful postselection and postselected measurements
both achieve the Heisenberg limit of quantum metrology, and show weak
measurement with thermal states only obtain classical Fisher information
(CFI) which increases with increasing temperature and achieves classical
enhanced scaling of $N^{2}$. Moreover, weak measurement with thermal states
has an advantage over that with coherent states or mixed states of the light
because generating these states with more large uncertainty are limited
under the current technology, but thermal states with more large uncertainly
are very easy to achieve with increasing temperature in nature, regardless
of thermal states of the light or the matter. ~~~~~\newline
~~~~~\newline
PACS numbers: 42.50.Wk, 42.65.Hw, 03.65.Ta
\end{abstract}

\maketitle

\section{ I. INTRODUCTION}

Quantum metrology is committed to enhancing measurement precision and
developing measurement techniques that give better precision than the same
measurement performed in a classical framework, therefore, it has aroused
substantial interest owing to its vital importance in physics and other
sciences \cite{Schnabel2010,Danilishin2012,Adhikari2014,Derevianko2011,Ludlow2015,Kolobov1999,Lugiato2002,Dowling2015,Aspelmeyer2014,Taylor2016}. By now, for the estimation of a parameter $\chi $ with a
pointer state that contains on average $N$ particles, a major way to
improving measurement precision is by utilizing non-classical resources,
such as quantum entanglement \cite{Bollinger1996,Walther2004,Afek2010} and squeezed states \cite{Goda2008,Grangier1987,Xiao1987,Treps2003}, which indicate that an
improved measurement precision can surpass the standard quantum limit ($%
\delta \chi \propto 1/\sqrt{N}$) or even achieves the Heisenberg scaling ($%
\delta \chi \propto 1/N$). However, the difficulty in generating highly
entangled states and fragility of such states is the open challenge to
enhance measurement precision beyond classical techniques in practical
applications. Moreover, whether quantum resource is essential for quantum
enhanced precision. Recently, by considering photon coupling with coherent
state pointer, the quantum Fisher information (QFI) can show a quantum
scaling of $N^{2}$ \cite{Zhang2015}, even without any quantum resources, and soon it was
experimentally verified via coherent light as a pointer \cite{Chen2018}. Meanwhile, for
the mixed states with modulating the power of coherent light, measurement
precision can experimentally attain Heisenberg scaling in weak measurement
\cite{G-Chen2018}. Thermal states are real classical states in nature, but they are not
considered as the pointers in the original weak measurement protocol because
their thermal fluctuations increase with temperature \cite{Aharonov1988}. Therefore, in most
weak measurement studies, the pure states (i.e., Gaussian states) are
generally considered to be the pointers. However, we have shown that by
using thermal states as the pointers can amplify weak measurement effect \cite{Li2015}.
These results motivate us to pursue innovative precision metrology schemes.

In this paper, we study quantum metrology of thermal states based on weak
measurement to bridge this gap. We show that in the displacement space
corresponding to imaginary weak values, at weak measurement limit the
maximal QFI after successful post-selection can attain the level of thermal
fluctuations, without surpassing total QFI, that weak measurement with
thermal states offer better precision relative to weak measurement with pure
states (i.e., Gaussian states), and that as the temperature increases, QFI
is also increased, thereby constantly improving the precision of parameter
estimation. On the other hand, for Kerr nonlinear interaction systems with
weak measurement, we surprisingly find that in the phase space successful
postselection and postselected measurement can beat the standard quantum
limit and achieve the Heisenberg limit of quantum metrology using classical
resources, i.e., thermal states, that it is shown that weak measurement with
thermal states only obtain classical Fisher information (CFI) increasing
when temperature increases, and it has an advantage over weak measurement
with coherent states or mixed states of the light because generating these
states with large uncertainty are limited under current technology, but
thermal states with large uncertainly are very easy to achieve with
increasing temperature in nature.

The structure of our paper is as follows. In Sec. II, we give a discussion
of quantum metrology of weak measurement with a thermal state pointer,
including precision measurement in the displacement space and the phase
space. In Sec. III, we obtain the conclusion of the work.

\section{II. Quantum metrology of weak measurement with thermal state pointer}

\subsection{1. Precision measurement in the displacement space}

\subsubsection{I. Weak measurement amplification model}

In the standard scenario of weak measurement \cite{Aharonov1988,Li2015,Simon2011,Li2014,Li-2015,Li15}, the
interaction Hamiltonian between the system and the pointer is (assuming $\hbar =1$)
\begin{equation}
\hat{H}=\hbar \chi (t)\hat{A}\otimes \hat{q},  \label{aa}
\end{equation}
where $A$ is a system observable, $q$ is the position observable of the
pointer and $A$ to be measured is usually a two-level system and the pointer
is a continuous system. $\chi (t)$ is a narrow pulse function with
interaction strengrh $\chi $. Suppose the initial system state is $|\psi
_{i}\rangle =\sum\limits_{i=1}^{2}c_{i}|a_{i}\rangle $ with $A|a_{i}\rangle
=a_{i}|a_{i}\rangle $, where $c_{1}=\cos \frac{\theta _{i}}{2}$ and $%
c_{2}=e^{i\varphi }\sin \frac{\theta _{i}}{2}$ with $0\leq \theta _{i}\leq
\pi $. Then we consider the initial pointer state as $\rho
_{th}(z)=(1-z)\sum\limits_{n=0}^{\infty }z^{n}|n\rangle _{m}\langle n|_{m}$
with $z=e^{-\hbar \omega _{m}/k_{B}T}$, where $k_{B}$ is the Boltzmann
constant and $T$ is the temperature.

The evolution of the total system after an interaction $U(t)=e^{-i\chi Aq}$
is given by
\begin{equation}
\rho (z)=e^{-i\chi Aq}|\psi _{i}\rangle \langle \psi _{i}|\otimes \rho
_{th}(z)e^{i\chi Aq}.  \label{ck}
\end{equation}

For Eq. (\ref{ck}), in momentum $p$ coordinate space we have
\begin{eqnarray}
\rho ^{p}(z) &=&\sum\limits_{i,j=1,n=0}^{2,\infty }\iint\limits_{-\infty
}^{\infty }\frac{(1-z)z^{n}}{2^{n}n!}_{i}c_{i}c_{j}^{\ast }|a_{i}\rangle
\langle a_{j}|G_{n}(p_{i})  \notag \\
&\times &G_{n}(p_{j}^{\prime })dpdp^{\prime }|p\rangle \langle p^{\prime }|,
\label{ch}
\end{eqnarray}
\begin{equation}
G_{n}(p_{i})=H_{n}(\sqrt{2}\sigma p_{i})\phi _{0}(p_{i}),  \label{kkji}
\end{equation}
where $\phi _{0}(p_{i})=(\frac{2\pi }{4\sigma ^{2}})^{-1/4}\exp (-\sigma
^{2}p_{i}^{2})$ with $p_{i}=p+a_{i}\chi $ and $\sigma $ is zero point
fluctuation, and $H_{n}$ is Hermite Polynomial.

Under this dynamics, each of the eigenstates $|a_{i}\rangle $ of the system
observable $A$ is entangled with the pointer state wavefunctions, which is
translated by the different $a_{i}\chi $ proportional to the eigenvalue $%
a_{i}$. When $|a_{i}-a_{j}|\chi $ is much larger than the width $\sqrt{\frac{
1+z}{1-z}}\frac{1}{2\sigma }$ of $\rho _{th}(z)$, this becomes a strong
measurement, meaning that the overlap
\begin{eqnarray}
O_{ij}:= &&(\frac{1+z}{1-z})^{-1/2}\sum\limits_{i\neq j}\int \exp [(4\sigma
^{2}p_{i}p_{j}z-2\sigma ^{2}(p_{i}^{2}  \notag \\
&+&p_{j}^{2})z^{2})/(1-z^{2})]\phi _{0}(p_{i})\phi _{0}(p_{j})dp
\label{ppoh}
\end{eqnarray}
between each pair of shifted wavefunctions is vanishingly small. So the
pointer state corresponding to different eigenvalues becomes completely
separated. However, when $|a_{i}-a_{j}|\chi $ is relatively small and the
wavefunctions are no longer well resolved, the measurement is said to be
weak \cite{Mello2010}.

When the postselected state of the measured system $|\psi _{a}\rangle
=\sum\limits_{i=1}^{2}c_{i}^{\prime }|a_{i}\rangle $ with $c_{1}^{\prime
}=\cos \frac{\theta _{f}}{2}$and $c_{2}^{\prime }=-\sin \frac{\theta _{f}}{2}
$ ($0\leq \theta _{f}\leq \pi $) is performed for the total system (\ref{ch}
), then $\rho ^{p}(z)$ reduces to
\begin{eqnarray}
\rho ^{p}(z) &=&\sum\limits_{i,j=1,n=0}^{2,\infty }\iint\limits_{-\infty
}^{\infty }c_{i}c_{i}^{\prime \ast }c_{j}^{\prime }c_{j}^{\ast }\frac{
(1-z)z^{n}}{2^{n}n!}G_{n}(p_{i})  \notag \\
&\times &G_{n}^{\ast }(p_{j}^{\prime })dpdp^{\prime }|p\rangle \langle
p^{\prime }|.  \label{ci}
\end{eqnarray}

After measuring the pointer in the $p$ basis, and using identity $\psi
(p)=\int\nolimits_{-\infty }^{\infty }\delta (p-p^{\prime })\psi (p^{\prime
})dp^{\prime }$, and Mehler's Hermite Polynomial Formula $%
\sum\limits_{n=0}^{\infty }\frac{H_{n}(x)H_{n}(y)}{n!}(\frac{w}{2}
)^{n}=(1-w^{2})^{-1/2}\exp [\frac{2xyw-(x^{2}+y^{2})w^{2}}{1-w^{2}}]$, the
probability distribution of Eq. (\ref{ci}) over $p$ becomes
\begin{align}
P_{wm}(p)& =\frac{1}{P_{a}}(2\pi \nabla ^{2})^{-1/2}[r^{2}\exp [-\frac{
(p+a_{1}\chi )^{2}}{2\nabla ^{2}}]  \notag \\
& -2rt(e^{-i\varphi }+e^{i\varphi })\exp [-\frac{(p+\frac{a_{2}-za_{1}}{1-z}
)^{2}}{4\nabla ^{2}}  \notag \\
& -\frac{(p+\frac{a_{1}-za_{2}}{1-z})^{2}}{4\nabla ^{2}}]+t^{2}\exp [-\frac{
(p+a_{2}\chi )^{2}}{2\nabla ^{2}}],  \label{hk}
\end{align}
where $r=\cos \frac{\theta _{i}}{2}\cos \frac{\theta _{f}}{2}$, $t=\sin
\frac{\theta _{i}}{2}\sin \frac{\theta _{f}}{2},\nabla ^{2}$ $=\frac{1+z}{%
1-z }\frac{1}{4\sigma ^{2}}$, and the probability of successful
postselection is
\begin{equation}
P_{a}=r^{2}+t^{2}-2rt\exp [-\frac{(a_{1}-a_{2})^{2}\chi ^{2}\Delta ^{2}}{2}
]\cos \varphi ,  \label{iiup}
\end{equation}
where $\Delta ^{2}=\frac{1+z}{1-z}\sigma ^{2}$.

After postselection measurement, in position $q$ coordinate space the
reduced state of Eq. (\ref{ck}) is given by
\begin{eqnarray}
\rho ^{q}(z) &=&\sum\limits_{i,j=1,n=0}^{2,\infty }\iint\limits_{-\infty
}^{\infty }c_{i}c_{i}^{\prime \ast }c_{j}^{\prime }c_{j}^{\ast }\frac{
(1-z)z^{n}}{2^{n}n!}G_{n}(q)  \notag \\
&\times&G_{n}(q^{\prime })e^{-i\chi q(a_{i}-a_{j})}dqdq^{\prime }|q\rangle
\langle q^{\prime }|,  \label{ou}
\end{eqnarray}
\begin{equation}
G_{n}(q)=H_{n}(\frac{q}{\sqrt{2}\sigma })\phi _{0}(q),  \label{ooiu}
\end{equation}
where $\phi _{0}(q)=(2\pi \sigma ^{2})^{1/4}\exp (-\frac{q^{2}}{4\sigma ^{2}}
)$.

After measuring the pointer in $q$ basis, the probability distribution of
Eq. (\ref{ou}) over $q$ is given by
\begin{eqnarray}
P_{wm}(q) &=&\frac{1}{P_{a}}(2\pi \Delta ^{2})^{-1/2}\exp [-\frac{q^{2}}{
2\Delta ^{2}}][r^{2}-rt  \notag \\
&\times &(\exp [i\varphi -i(a_{2}-a_{1})\chi q]+\exp [i(a_{2}  \notag \\
&-&a_{1})\chi q-i\varphi ])+t^{2}].  \label{yyi}
\end{eqnarray}

When the postselection fails (with probability $P_{r}=1-P_{a}$), namely, the
failing postselected state of the measured system $|\psi _{r}\rangle
=\sum\limits_{i=1}^{2}c_{i}^{\prime }|a_{i}\rangle $ with $c_{1}^{\prime
}=\sin \frac{\theta _{f}}{2}$and $c_{2}^{\prime }=\cos \frac{\theta _{f}}{2}$
is performed for the total system (\ref{ck}), we calculate carefully and
obtain the probalility distribution of failing postselection in $p$ basis
\begin{align}
P_{wm}(p)&= \frac{1}{P_{r}}(2\pi \nabla ^{2})^{-1/2}[r^{\prime 2}\exp [-
\frac{(p+a_{1}\chi )^{2}}{2\nabla ^{2}}]  \notag \\
&+ 2r^{\prime }t^{\prime }(e^{-i\varphi }+e^{i\varphi })\exp [-\frac{(p+
\frac{a_{2}-za_{1}}{1-z})^{2}}{4\nabla ^{2}}  \notag \\
&- \frac{(p+\frac{a_{1}-za_{2}}{1-z})^{2}}{4\nabla ^{2}}]+t^{\prime 2}\exp
[- \frac{(p+a_{2}\chi )^{2}}{2\nabla ^{2}}],  \label{uupo}
\end{align}
and the probalility distribution of failing postselection in $q$ basis
\begin{align}
P_{wm}(q)& =\frac{1}{P_{r}}(2\pi \Delta ^{2})^{-1/2}\exp [-\frac{q^{2}}{
2\Delta ^{2}}][r^{\prime 2}+r^{\prime }t^{\prime }  \notag \\
& \times (\exp [i\varphi -i(a_{2}-a_{1})\chi q]+\exp [i(a_{2}  \notag \\
& -a_{1})\chi q-i\varphi ])+t^{\prime 2}],  \label{oop}
\end{align}
where $r^{\prime }=\cos \frac{\theta _{i}}{2}\sin \frac{\theta _{f}}{2}$and $%
t^{\prime }=\sin \frac{\theta _{i}}{2}\cos \frac{\theta _{f}}{2}.$

\subsubsection{II. Metric}

In the precision metrology, a parameter estimation interested can be given
by the Fisher information (FI) \cite{Helstrom1976}, and it is functional on such conditional
probability distributions, and is defined as follows
\begin{equation}
F_{\chi }[P(s|\chi )]=\int_{-\infty }^{\infty }\frac{(\partial _{x}P(s|\chi
))^{2}}{P(s|\chi )}ds,  \label{kll}
\end{equation}
where $s$ represents $p$ or $q$. The sensitive estimate of an unknown
parameter is given by the observed statistics \cite{Cramer1946}, i.e., Cramer-Rao bound
limits
\begin{equation}
Var(\chi )\geq \frac{1}{NF_{\chi }},  \label{opp}
\end{equation}
where $Var()$ is a variance expression and $N$ is the number of independent
trials.

If the postselected weak measurement is applied to the precision metrology
of a parameter estimation, the whole process is called weak measurement
amplification strategy (i.e., WMA strategy). Conversely, the standard
strategy is when there is no weak measurement, and it refers to the
benchmark measurement strategy completely ignoring the degree of freedom of
the system. For Eq. (\ref{ch}), one traces over the degree of freedom of the
system and measures the particle in $p$ basis to give
\begin{equation}
P_{\mathbf{std}}^{joint}(p,z)=\sum\limits_{i=1}^{2}|c_{i}|^{2}\phi
(p+a_{i}\chi ),  \label{op}
\end{equation}
where $\phi (p+a_{i}\chi )=(2\pi \nabla ^{2})^{-1/2}\exp [-(p+a_{i}\chi
)^{2}/(2\nabla ^{2})]$.

Substitute (\ref{op}) into the FI fomula (\ref{kll}) and applying the
Cauchy-Schwarz inequality, the quantum Fisher information (QFI) of the joint
system state before postselection in the momentum $p$ coordinate space is
\begin{align}
Q_{j}(p)& =\nabla ^{-4}\int_{-\infty }^{\infty }\frac{[\sum%
\limits_{i=1}^{2}|c_{i}|^{2}\phi (p+a_{i}\chi )(p+a_{i}\chi )a_{i}]^{2}}{P_{
\mathbf{std}}^{joint}(p|\chi )}dp  \notag \\
& \leq \nabla ^{-4}\int_{-\infty }^{\infty
}\sum\limits_{i=1}^{2}|c_{i}|^{2}a_{i}^{2}\phi (p+a_{i}\chi )(p+a_{i}\chi
)^{2}dp  \notag \\
& =\nabla ^{-2}\sum\limits_{i=1}^{2}|c_{i}|^{2}a_{i}^{2}  \notag \\
& \leq |a_{i}|_{\max }^{2}\nabla ^{-2}.  \label{ioo}
\end{align}

We let $F_{\chi }(P_{\mathbf{std}}(p,z))$ take the maximum value of (\ref%
{ioo}), i.e., $F_{\chi }(P_{\mathbf{std}}(p,z))=|a_{i}|_{\max }^{2}\nabla
^{-2}$. However, when $z=0$,
\begin{equation}
F_{\chi }(P_{\mathbf{std}}(p,z=0))=|a_{i}|_{\max }^{2}(\frac{1}{4\sigma ^{2}}
)^{-1},  \label{uuu}
\end{equation}
which shows the highest FI of pure Gaussian states in a standard strategy.
It can be seen from Eq. (\ref{uuu}) that in a standard strategy using pure
Gaussian states give the higher estimate of an unkown parameter $\chi $ than
using thermal states. Therefore, the use of thermal states in weak
measurement has no advantage over the use of pure states in terms of
conventional measurement. Here we use QFI of pure states as the benchmark in
the standard strategy.

In the WMA strategy, the QFI can be divided into three parts, and $%
F_{Q}^{tot}=P_{a}F_{Q}^{a}+P_{r}F_{Q}^{r}+F_{Q}^{p}$ (see QFI derivation of
\cite{Zhang2015}), where $F_{Q}^{a}$ and $F_{Q}^{r}$ denote the QFI of successful
postselection (accepted information) and failing postselection (rejected
information) for weak measurement, respectively, and $F_{Q}^{p}$ is
classical FI for projective measurement. Since $F_{Q}^{tot}(p)\leq Q_{j}(p)$
\cite{Zhang2015}, and in the momentum $p$ coordinate space, $Q_{j}(p)\leq F_{\chi }(P_{
\mathbf{std}}(p,z)\leq F_{\chi }(P_{\mathbf{std}}(p,z=0))$. Therefore, $%
F_{Q}^{tot}(p)\leq Q_{j}(p)\leq F_{\chi }(P_{\mathbf{std}}(p,z=0))$.

Substitute (\ref{yyi}) and (\ref{oop}) into the FI fomula (\ref{kll}),
respectively, in the position $q$ coordinate space,
\begin{align}
F_{Q}^{a}(q)& =\frac{-4r^{2}t^{2}}{P_{a}^{2}}\exp [-(a_{2}-a_{1})^{2}\chi
^{2}\Delta ^{2}]\cos ^{2}\varphi \Delta ^{4}  \notag \\
& \times (a_{2}-a_{1})^{4}\chi ^{2}+\frac{r^{2}t^{2}}{P_{a}}
(a_{2}-a_{1})^{2}(2\pi \Delta ^{2})^{-1/2}  \notag \\
& \times \int_{-\infty }^{\infty }q^{2}\exp (-\frac{q^{2}}{2\Delta ^{2}}
)(4-(\exp [i(a_{2}-a_{1})\chi q  \notag \\
& -i\varphi )]+\exp [i\varphi -i(a_{2}-a_{1})\chi q)])^{2})dq/[r^{2}  \notag
\\
& +t^{2}-rt(\exp [i(a_{2}-a_{1})\chi q-i\varphi )]+\exp [i\varphi  \notag \\
& -i(a_{2}-a_{1})\chi q])],  \label{uuy}
\end{align}
and
\begin{align}
F_{Q}^{r}(q)& =\frac{-4r^{2}t^{2}}{P_{r}^{2}}\exp [-(a_{2}-a_{1})^{2}\chi
^{2}\Delta ^{2}]\cos ^{2}\varphi \Delta ^{4}  \notag \\
& \times (a_{2}-a_{1})^{4}\chi ^{2}+\frac{r^{2}t^{2}}{P_{r}}
(a_{2}-a_{1})^{2}(2\pi \Delta ^{2})^{-1/2}  \notag \\
& \times \int_{-\infty }^{\infty }q^{2}\exp (-\frac{q^{2}}{2\Delta ^{2}}
)(4-(\exp [i(a_{2}-a_{1})\chi q  \notag \\
& -i\varphi )]+\exp [i\varphi -i(a_{2}-a_{1})\chi q)])^{2})dq/[r^{\prime 2}
\notag \\
& +t^{\prime 2}+rt(\exp [i(a_{2}-a_{1})\chi q-i\varphi )]+\exp [i\varphi
\notag \\
& -i(a_{2}-a_{1})\chi q])].  \label{uuyi}
\end{align}
Moreover,
\begin{eqnarray}
F_{Q}^{P} &=&\frac{P_{a}^{\prime ^{2}}}{P_{a}}+\frac{P_{r}^{\prime ^{2}}}{
P_{r}}  \notag \\
&=&(\frac{4r^{2}t^{2}}{P_{a}}+\frac{4r^{2}t^{2}}{P_{r}})\exp
[-(a_{2}-a_{1})^{2}\chi ^{2}\Delta ^{2}]  \notag \\
&\times &\cos ^{2}\varphi \Delta ^{4}(a_{2}-a_{1})^{4}\chi ^{2}.
\end{eqnarray}
Therefore,
\begin{align}
& F_{Q}^{tot}(q)=r^{2}t^{2}(a_{2}-a_{1})^{2}(2\pi \Delta
^{2})^{-1/2}\int_{-\infty }^{\infty }\exp (-\frac{q^{2}}{2\Delta ^{2}})
\notag \\
& \times q^{2}(4-(\exp [i(a_{2}-a_{1})\chi q-i\varphi )]+\exp [i\varphi
\notag \\
& -i(a_{2}-a_{1})\chi q)])^{2})dq(1/[r^{2}+t^{2}-rt(\exp [i(a_{2}  \notag \\
& -a_{1})\chi q-i\varphi )]+\exp [i\varphi -i(a_{2}-a_{1})\chi q])]  \notag
\\
& +1/[r^{\prime 2}+t^{\prime 2}+rt(\exp [i(a_{2}-a_{1})\chi q-i\varphi )]
\notag \\
& +\exp [i\varphi -i(a_{2}-a_{1})\chi q])])  \notag \\
& \leq 4rt(a_{2}-a_{1})^{2}(2\pi \Delta ^{2})^{-1/2}\int_{-\infty }^{\infty
}q^{2}\exp (-\frac{q^{2}}{2\Delta ^{2}})  \notag \\
& =4rt(a_{2}-a_{1})^{2}\Delta ^{2},  \label{iiyi}
\end{align}
and the equality holds up if and only if $\cos \frac{\theta _{i}}{2}=\sin
\frac{\theta _{i}}{2}$and $\cos \frac{\theta _{f}}{2}=\sin \frac{\theta _{f}
}{2}$. Set the average number of thermal states $\langle n\rangle =\frac{z}{
1-z}$, the maximal account of $F_{Q}^{tot}(q)$ can be rewritten as $%
(a_{2}-a_{1})^{2}(2\langle n\rangle +1)\sigma ^{2}$. Thus, the scaling is at
the standard quantum limit. However, in the position $q$ coordinate space,
there is no information for the joint state (\ref{ck}), i.e., $Q_{j}(q)=0$.
Indicate that the QFI of Eq. (\ref{uuy})-Eq. (\ref{iiyi}) in the position
space originate in postselected measurement, that is, QFI in the position
space can be generated as long as projective measurement occurs.

However, our interesting attention in this paper is the Cramer-Rao bound for
the WMA strategy, and how it compares to that for the standard strategy. In
the limit of $N\rightarrow \infty $, their ratio is equal to $%
P_{a}F_{Q}^{a}(p)$ and $P_{a}F_{Q}^{a}(q)$ to $F_{\chi }(P_{\mathbf{std}
}(p,z=0))$, respectively. The formers are corrected by the probability of
successful postselection.

\subsubsection{III. Ideal detector}

Here we consider a stable and ideal detector (i.e., without technical
imperfection). In the WMA strategy, if choosing to measure in momentum space
or position space, one will obtain the displacement proportional to the real
part of weak values or one proportional to the imaginary part of weak
values, respectively. The corresponding conditional probability distribution
is given by (\ref{hk}) or (\ref{yyi}). Then taking a ratio of the WMA
strategy to the standard strategy, for (\ref{hk}) and (\ref{uuu}) we give

\begin{eqnarray}
&&\frac{P_{a}F_{Q}^{a}[P_{wm}(p)]}{F_{\chi }(P_{\mathbf{std}}(p,z=0))}
\notag \\
&=&P_{a}\int_{-{\infty }}^{{\infty }}\frac{(\partial _{\chi }P_{wm}(p))^{2}}{
P_{wm}(p)}dp/[|a_{i}|_{\max }^{2}(\frac{1}{4\sigma ^{2}})^{-1}].
\label{yyiu}
\end{eqnarray}
The numberator of (\ref{yyiu}) is QFI for the momentum's displacement
proportional to real weak values. Note that is no larger than QFI of the
joint system state (\ref{ci}) without postselection \cite{Zhang2015}, i.e., $
Q_{j}(p)=|a_{i}|_{\max }^{2}\nabla ^{-2}$. However, in weak measurement
limit, i.e., $\chi \sigma \rightarrow 0$, $P_{a}\int_{-{\infty }}^{{\infty }%
} \frac{(\partial _{\chi }P_{wm}(p))^{2}}{P_{wm}(p)}dp$ is equal to $%
(ra_{1}-ta_{2})^{2}\nabla ^{-2}$. Therefore, $\frac{%
P_{a}F_{Q}^{a}[P_{wm}(p)] }{F_{\chi }(P_{\mathbf{std}}(p,z=0))}\leq \frac{1-z%
}{1+z}$, and the equality holds up if and only if $\cos \frac{\theta _{i}}{2}%
=\sin \frac{\theta _{i}}{2 }$and $\cos \frac{\theta _{f}}{2}=\sin \frac{%
\theta _{f}}{2}$. It is clear that QFI for real weak values space using
thermal state pointer can be no advantage for the purpose of estimating $%
\chi $.

Multiplying Eq. (\ref{iiup}) by Eq. (\ref{uuy}), I further give
\begin{eqnarray}
P_{a}F_{Q}^{a}(q) &\leq &\frac{-4r^{2}t^{2}\Delta ^{4}(a_{2}-a_{1})^{4}\chi
^{2}}{P_{a}}\exp [-(a_{2}  \notag \\
&-&a_{1})^{2}\chi ^{2}\Delta ^{2}]\cos ^{2}\varphi +2rt\Delta
^{2}(a_{2}-a_{1})^{2}  \notag \\
&\times &(1+\exp [-\frac{(a_{2}-a_{1})^{2}\chi ^{2}\Delta ^{2}}{2}][1-(a_{2}
\notag \\
&-&a_{1})^{2}\chi ^{2}\Delta ^{2}]\cos \varphi ),  \label{iiiup}
\end{eqnarray}
and the equality holds up if and only if $\cos \frac{\theta _{i}}{2}=\sin
\frac{\theta _{i}}{2}$and $\cos \frac{\theta _{f}}{2}=\sin \frac{\theta _{f}
}{2}$. Here we only consider WMA strategy, $F_{Q}^{tot}(q)=P_{a}F_{Q}^{a}(q)$
. In the weak measurement limit, defined as $\chi \Delta \rightarrow 0$, $%
P_{a}F_{Q}^{a}(q)=(a_{2}-a_{1})^{2}\Delta ^{2}(1+\cos \varphi )$, and the
maximum value is $(a_{2}-a_{1})^{2}\Delta ^{2}$ when $\varphi =0\ $or $2\pi $
, i.e., $|\psi _{i}\rangle $ and $|\psi _{a}\rangle $ are completely
orthogonal, but $P_{r}F_{Q}^{r}=0$ and $F_{Q}^{P}=0$. In the strong
measurement limit, when $\chi \Delta \gg 1$, $%
P_{a}F_{Q}^{a}(q)=(a_{2}-a_{1})^{2}\Delta ^{2}/2$, $%
P_{r}F_{Q}^{r}(q)=(a_{2}-a_{1})^{2}\Delta ^{2}/2$, and $F_{Q}^{P}=0$. These
results are discussed in the context of imaginary weak values space in weak
measurement. It can be seen that when the uncertainty of the pointer state $%
\Delta $ is fixed, the measurement precision in weak measurement limit is no
better than that in strong measurement limit. However, when the parameter $%
\chi $ is fixed, the measurement precision is always better because using a
pointer state with larger $\Delta $ is determined by high temperature. Thus
QFIs are all proportional to $(a_{2}-a_{1})^{2}\Delta ^{2}$.

However, for (\ref{iiiup}) and (\ref{uuu}) we have
\begin{align}
& \frac{P_{a}F_{Q}^{a}[P_{wm}(q)]}{F_{\chi }(P_{\mathbf{std}}(p,z=0))}
\notag \\
& \leq \frac{(a_{2}-a_{1})^{2}}{|a_{i}|_{\max }^{2}}(\frac{1+z}{1-z})[-\frac{
r^{2}t^{2}}{P_{a}}\exp (-k^{2})k^{2}\cos ^{2}\varphi  \notag \\
& +\frac{1}{2}rt(1+\exp (-k^{2}/2)(1-k^{2})\cos \varphi )],  \label{uuyui}
\end{align}
where $k^{2}=(a_{2}-a_{1})^{2}\chi ^{2}\Delta ^{2}$, and the equality holds
up if and only if $\cos \frac{\theta _{i}}{2}=\sin \frac{\theta _{i}}{2}$
and $\cos \frac{\theta _{f}}{2}=\sin \frac{\theta _{f}}{2}$. The numberator
of ( \ref{uuyui}) is QFI for the position's displacement proportional to the
imaginary weak values.

Suppose the eigenvalues $a_{1}=1$, $a_{2}=-1$, implying that $|a_{i}|_{\max
}=1$, and $\theta _{i}=\theta _{f}=\pi /2$. In weak measurement regime,
i.e., $k\ll 1$ and $\varphi \ll 1$, for example, $k^{2}=0.0005$ with $z=0$, $%
\varphi =0.05$, $\frac{P_{a}F_{Q}^{a}[P_{wm}(q)]}{F_{\chi }(P_{\mathbf{std}
}(p,z=0))}=0.8327$ and $k^{2}=0.0095$ with $z=0$, $\varphi =0.05$, $\frac{
P_{a}F_{Q}^{a}[P_{wm}(q)]}{F_{\chi }(P_{\mathbf{std}}(p,z=0))}=3.9478$.
Moreover, when $\varphi =\pi /2$, $\frac{P_{a}F_{Q}^{a}[P_{wm}(q)]}{F_{\chi
}(P_{\mathbf{std}}(p,z=0))}=\frac{1+z}{1-z}/2$. These results suggest that
as $z$ grows, the ratio of (\ref{uuyui}) can exceed $1$. In other words, by
adjusting the temperature $T$, we can give a better estimate of an unknown
parameter $\chi $. Therefore, the result breaks the inequality constraint in \cite{Knee2014,Knee2015,Tanaka2013}
\begin{equation}
{P_{\mathbf{postselection}}}{F_{\mathbf{weak\kern1ptvalue}}}\leq {F_{\mathbf{standard}},}  \label{hnyt}
\end{equation}
where is the postselection success probability. It is obvious that
postselected weak measurement using thermal state pointer, corresponding to
the displacement proportional to imaginary weak values, can increase the
measurement precision.

\subsection{2. Precision measurement in the phase space}

We now consider a scenario that particle-number distribution. The initial
state of the quantum system $|\psi _{i}\rangle $ and one of the quantum
pointer $\rho _{th}(z)$ are both the same as before, while let the
observable of the pointer be $\hat{n}$. Thus in the Hamiltonian of Eq. (\ref%
{aa}) we only use $\hat{n}$ instead of $\hat{q}$, the other is unchanged.

After the interaction $U(t)=e^{-i\chi An}$ for the initial state of the
total system, its time evolution is given by

\begin{equation}
\rho (z)=e^{-i\chi An}|\psi _{i}\rangle \langle \psi _{i}|\otimes \rho
_{th}(z)e^{i\chi An}.  \label{ckk}
\end{equation}

Using QFI of the pure state $|\lambda _{m}\rangle $: $F_{Q,m}=4[\langle
\lambda _{m}^{\prime }|\lambda _{m}^{\prime }\rangle -|\langle \lambda
_{m}|\lambda _{m}^{\prime }\rangle |^{2}]$ with $\lambda _{m}^{\prime
}=\partial \lambda _{m}/\partial \chi $, $|\lambda _{m}^{\prime }\rangle
=\partial |\lambda _{m}\rangle /\partial \chi $, and the convexity of the
QFI of the mixed state: $Q_{j}\leq \sum\limits_{m}\lambda _{m}F_{Q,m}$, we
can obtain QFI of the joint state (\ref{ckk})
\begin{equation}
Q_{j}=(a_{1}-a_{2})^{2}(2\langle n\rangle ^{2}+\langle n\rangle )\sin
^{2}\theta _{i},  \label{uu}
\end{equation}
where $\langle n\rangle =\frac{z}{1-z}$. $Q_{j}$ is the maximum amount of
information when $\theta _{i}=\pi /2$, and proportional to quantum scaling ($%
\sim \langle n\rangle ^{2}$). From the expression of Eq. (\ref{uu}), it can
be seen that $\theta _{i}=0$, $\pi $ will never provide a better than
classical scaling. These correspond to two cases, namely, for the
eigenstates of $A$: $|\psi _{i}\rangle =|a_{1}\rangle $ or $|a_{2}\rangle $,
$P_{a}F_{Q}^{a}=0$, $P_{r}F_{Q}^{r}=0$, and $F_{Q}^{P}=0$. Thus, $%
F_{Q}^{tot}=0$.

When the postselected state of the measured system $|\psi _{a}\rangle
=\sum\limits_{i=1}^{2}c_{i}^{\prime }|a_{i}\rangle $ with $c_{1}^{\prime
}=\cos \frac{\theta _{f}}{2}$and $c_{2}^{\prime }=-\sin \frac{\theta _{f}}{2}
$is made for the total system (\ref{ckk}), the reduced state of the pointer
is $\rho _{a}^{n}(z)=\sum\limits_{n=0}^{\infty }\Theta _{a}|n\rangle
_{m}\langle n|_{m}$ (unnormalized), where
\begin{eqnarray}
\Theta _{a} &=&(1-z)z^{n}[r^{2}+t^{2}-rt(e^{i[\varphi +\chi (a_{1}-a_{2})n]}
\notag \\
&+&e^{-i[\varphi +\chi (a_{1}-a_{2})n]})].  \label{lluj}
\end{eqnarray}
Using identity $\frac{1}{1-z}=\sum\limits_{n=0}^{\infty }z^{n}$, the
probability of successful postselection is

\begin{eqnarray}
P_{a} &=&r^{2}+t^{2}-rt[e^{-i\varphi }(1-z)/(1-ze^{-i\chi (a_{1}-a_{2})})
\notag \\
&+&e^{i\varphi }(1-z)/(1-ze^{i\chi (a_{1}-a_{2})})].  \label{oo}
\end{eqnarray}
Thus, we can give the normalized state by
\begin{equation}
\rho _{a}^{n}(z)=\sum\limits_{n=0}^{\infty }\frac{\Theta _{a}}{P_{a}}
|n\rangle _{m}\langle n|_{m}.  \label{kk}
\end{equation}

Since $\rho _{a}^{n}(z)$ is a mixed state, and satisfies diagonalization
form as $\rho =\sum\limits_{m}\lambda _{m}|\lambda _{m}\rangle \langle
\lambda _{m}|$, where \{$|\lambda _{m}\rangle $\} forms an orthogonality and
complete basis, with $\lambda _{m}$ being the weight of $|\lambda
_{m}\rangle $. According to the well-known formula, the QFI of the mixed
state with $\lambda _{m}\neq 0$ is given by (see QFI derivation of \cite{Knysh2011,Liu2014,Zhang2013})
\begin{eqnarray}
F_{Q} &=&\sum\limits_{m}\frac{(\lambda _{m}^{\prime })^{2}}{\lambda _{m}}
+\sum\limits_{m}\lambda _{m}F_{Q,m}-\sum\limits_{m\neq n}\frac{8\lambda
_{m}\lambda _{n}}{\lambda _{m}+\lambda _{n}}  \notag \\
&\times &|\langle \lambda _{m}|\lambda _{n}^{\prime }\rangle |^{2}.
\label{nm}
\end{eqnarray}
The first term is the classical Fisher information for the probability
distribution $P(m|\chi )=\lambda _{m}(\chi )$. The second term is weighted
average over the QFI $F_{Q,m}=4[\langle \lambda _{m}^{\prime }|\lambda
_{m}^{\prime }\rangle -|\langle \lambda _{m}|\lambda _{m}^{\prime }\rangle
|^{2}]$ for each pure state in the subset \{$|\lambda _{m}\rangle $\}, with $%
\lambda _{m}\neq 0$. The last term reduces the QFI and hence the estimation
precision below the pure state case.

Substitute (\ref{kk}) into the QFI fomula (\ref{nm}), we have
\begin{eqnarray}
F_{Q}^{a} &=&\sum\limits_{n=0}^{\infty }(\frac{\Theta _{a}^{\prime 2}}{
\Theta _{a}P_{a}}-\frac{2\Theta _{a}^{\prime }P_{a}^{\prime }}{P_{a}^{2}}+
\frac{\Theta _{a}P_{a}^{\prime ^{2}}}{P_{a}^{3}})  \notag \\
&=&\frac{1}{P_{a}}\sum\limits_{n=0}^{\infty }\frac{\Theta _{a}^{\prime 2}}{
\Theta _{a}}-\frac{P_{a}^{\prime ^{2}}}{P_{a}^{2}},  \label{gg}
\end{eqnarray}
where $\Theta _{a}^{\prime }=\partial \Theta _{a}/\partial \chi $, $%
P_{a}^{\prime }=\partial P_{a}/\partial \chi $, and $P_{a}^{\prime
}=\sum\limits_{n=0}^{\infty }\Theta _{a}^{\prime }$. It can be easily seen
that the FI in Eq. (\ref{gg}) is not QFI, but is described by the classical
Fisher information (CFI) for the probability distribution $\frac{\Theta _{a}
}{P_{a}}$ in Eq. (\ref{nm}).

When the postselection fails (with probability $P_{r}=1-P_{a}$), namely, the
failing postselected state of the measured system $|\psi _{r}\rangle
=\sum\limits_{i=1}^{2}c_{i}^{\prime }|a_{i}\rangle $ with $c_{1}^{\prime
}=\sin \frac{\theta _{f}}{2}$ and $c_{2}^{\prime }=\cos \frac{\theta _{f}}{2}
$ is performed for the total system (\ref{ckk}), the reduced state of the
pointer (normalized) is given by
\begin{equation}
\rho _{r}^{n}(z)=\sum\limits_{n=0}^{\infty }\frac{\Theta _{r}}{P_{r}}
|n\rangle _{m}\langle n|_{m},  \label{ggg}
\end{equation}
where
\begin{eqnarray}
\Theta _{r} &=&(1-z)z^{n}[r^{\prime 2}+t^{\prime 2}+rt(e^{i[\varphi +\chi
(a_{1}-a_{2})n]}  \notag \\
&+&e^{-i[\varphi +\chi (a_{1}-a_{2})n]})]  \label{kkju}
\end{eqnarray}
with $r^{\prime }=\cos \frac{\theta _{i}}{2}\sin \frac{\theta _{f}}{2}$and $%
t^{\prime }=\sin \frac{\theta _{i}}{2}\cos \frac{\theta _{f}}{2}$. Note that
the failing postselection is not considered in the original scheme and often
ignored in experiments. Hence, the CFI of the failing postselection is given
by

\begin{eqnarray}
F_{Q}^{r} &=&\sum\limits_{n=0}^{\infty }(\frac{\Theta _{r}^{\prime 2}}{
\Theta _{r}P_{r}}-\frac{2\Theta _{r}^{\prime }P_{r}^{\prime }}{P_{r}^{2}}+
\frac{\Theta _{r}P_{r}^{\prime ^{2}}}{P_{r}^{3}})  \notag \\
&=&\frac{1}{P_{r}}\sum\limits_{n=0}^{\infty }\frac{\Theta _{r}^{\prime 2}}{
\Theta _{r}}-\frac{P_{r}^{\prime ^{2}}}{P_{r}^{2}},  \label{gh}
\end{eqnarray}
where $\Theta _{r}^{\prime }=\partial \Theta _{r}/\partial \chi $, $%
P_{r}^{\prime }=\partial P_{r}/\partial \chi $, and $P_{r}^{\prime
}=\sum\limits_{n=0}^{\infty }\Theta _{r}^{\prime }$.

Obviously, the CFI after the projective measurement is given by \cite{Zhang2015}
\begin{equation}
F_{Q}^{P}=\frac{P_{a}^{\prime ^{2}}}{P_{a}}+\frac{P_{r}^{\prime ^{2}}}{P_{r}}
,  \label{fv}
\end{equation}
As $\chi \rightarrow 0$, $F_{Q}^{P}=\frac{4(a_{1}-a_{2})^{2}r^{2}t^{2}\sin
^{2}\varphi }{(1-r^{2}-t^{2}+2rt\cos \varphi )(r^{2}+t^{2}-2rt\cos \varphi )}
\langle n\rangle ^{2}$, implying that $F_{Q}^{P}$ can reach the limit of QFI
of the joint state $Q_{j}$ ($\sim $ $\langle n\rangle ^{2}$). Set $\theta
_{i}=\theta _{f}=\frac{\pi }{2}$, and we find that
\begin{equation}
F_{Q}^{P}=(a_{1}-a_{2})^{2}\langle n\rangle ^{2},  \label{ooopp}
\end{equation}
Moreover, in this same situation, the CFIs for both the successful and
failed postselection are $P_{a}F_{Q}^{a}=(a_{2}-a_{1})^{2}\sin ^{2}\varphi
(\langle n\rangle ^{2}+\langle n\rangle )/(2-2\cos \varphi )$ and $%
P_{r}F_{Q}^{r}=(a_{2}-a_{1})^{2}\sin ^{2}\varphi (\langle n\rangle
^{2}+\langle n\rangle )/(2+2\cos \varphi )$. This shows that $P_{a}F_{Q}^{a}$
can achieve its maximal amount
\begin{equation}
P_{a}F_{Q}^{a}=(a_{2}-a_{1})^{2}(\langle n\rangle ^{2}+\langle n\rangle )
\label{ooimk}
\end{equation}
when $\varphi \rightarrow 0$ or $2\pi $, i.e., $|\psi _{i}\rangle $ and $%
|\psi _{a}\rangle $ are completely orthogonal, but $P_{r}F_{Q}^{r}=0$. As
our calculation shows, whether the successful postselection or the
projective measurement (postselected measurement), their FIs indeed scales
at the Heisenberg limit ($F_{Q}\sim \langle n\rangle ^{2}$).

Hence,
\begin{eqnarray}
F_{Q}^{tot} &=&\sum\limits_{n=0}^{\infty }(\frac{\Theta _{d}^{\prime 2}}{
\Theta _{d}}+\frac{\Theta _{r}^{\prime 2}}{\Theta _{r}})  \notag \\
&\leq &4rt(a_{1}-a_{2})^{2}\frac{z(1+z)}{(1-z)^{2}}  \notag \\
&=&(a_{1}-a_{2})^{2}(2\langle n\rangle ^{2}+\langle n\rangle )\sin \theta
_{i}\sin \theta _{f},  \label{oouy}
\end{eqnarray}
and the equality holds up if and only if $\cos \frac{\theta _{i}}{2}=\sin
\frac{\theta _{i}}{2}$and $\cos \frac{\theta _{f}}{2}=\sin \frac{\theta _{f}
}{2}$. We find that $F_{Q}^{tot}\leq Q_{j}$, and equality holds up when $%
\theta _{i}=\theta _{f}$. The maximal $F_{Q}^{tot}\ $and $Q_{j}$ are found
for $\theta _{i}=\theta _{f}=\pi /2$, and equal to $(a_{1}-a_{2})^{2}(2
\langle n\rangle ^{2}+\langle n\rangle )$. The results show that the
Heisenberg scaling arises in weak measurement using thermal state as a
pointer.

\section{ Conclusion}

In summary, by considering thermal states as the pointers in weak
measurement, we have concluded that in the displacement space the QFI
corresponding to real weak values is no advantage for the precision of
parameter estimation. For imaginary weak values, we have shown that in the
case of weak coupling ($\chi \rightarrow 0$) the maximal QFI after
successful postselction can achieve the level of thermal fluctuations ($%
F_{Q}\sim \frac{1+z}{1-z}\sigma ^{2}$), and as the temperature increases,
QFI is also increased, thereby improving the precision of parameter
estimation, in sharp contrast with QFI using Gaussian states (i.e., pure
states) as the pointers in the standard strategy which indicates that QFI
only achieve the level of zero point fluctuations ($F_{Q}\sim \sigma ^{2}$).

In the phase space, however, our calculations show that not only successful
postselection but postselected measurement itself only contain useful CFI
when weak measurement use only classical resources, i.e., thermal states,
and in weak measurement limit ($\chi \rightarrow 0$) their CFI can both
attain the Heisenberg scaling ($F_{Q}\sim \langle n\rangle ^{2},\langle
n\rangle =\frac{z}{1-z}$) for the precision of parameter estimation.
Obviously, weak measurement using thermal state pointer in phase space can
yield calssical-enhanced precision. As the temperature increases, CFI is
futher increased. Thus, the measurement precision of the Heisenberg limit
can be much larger than that of the classical measurement method. It is a
known fact that thermal states are easy to be prepared under current
experimental conditions. Our work provides a way to realize Heisenberg
scaling precision, regardless of utilizing the light or the matter as a
pointer.

\section{ACKNOWLEDGMENT}

This work was supported by the Natural Science Foundation of Shaanxi Province (Grant No. 2018JQ1056), the Doctoral Scientific Research Foundation of Yan'an University (Grant No. YDBK2016-04) and Youth Foundation of Yan'an University (Grant No. YDQ2017-09). \newline

\end{document}